\documentclass[prd,aps,letterpaper,twocolumn,superscriptaddress,preprintnumbers,nofootinbib,floatfix]{revtex4}
\pdfoutput=1

\usepackage[pdftex]{graphicx}
\usepackage{amsmath}
\usepackage{amsfonts}
\usepackage{amssymb}
\usepackage{rotating}
\usepackage{subfigure}
\usepackage{paralist} 
\usepackage{verbatim}
\usepackage{color}
\usepackage{soul} 
\usepackage{appendix}

\usepackage{natbib}
\usepackage{epsfig,hyperref}
\usepackage{array}
\usepackage{threeparttable}

\usepackage[utf8]{inputenc}
\usepackage[vietnamese,english]{babel}
\usepackage{tabularx}
\newcolumntype{C}{>{\centering\arraybackslash}X}

\usepackage{mathpazo} 
\usepackage{tgpagella} 

\newcommand{\na}{New Astronomy}
\newcommand{\mnras}{Monthly Notices of the Royal Astronomical Society}
\newcommand{\apjl}{Astrophysical Journal Letters}

\newcommand{\aap}{Astronomy \& Astrophysics}
\newcommand{\jcap}{J. Cosmol. Astropart. Phys.}
\newcommand{\procspie}{{Proc.~SPIE}}
\newcommand{\Bell}{\boldsymbol{\ell}}
\newcommand{\bL}{\boldsymbol{L}}

\newcommand{\bll}{{\bf L}}
\newcommand{\dotfac}[1]{({\bll} \cdot {\bf \Bell}_{#1})}
\newcommand{\planck}{{\it{Planck~}}}
\newcommand{\T}{{\it{T}}}

\newcommand{\physrep}{Phys.~Rep.}

\usepackage{color}
\definecolor{orange}{rgb}{1,0.3,0}

\begin{document}

\title{Measuring the Small-Scale Matter Power Spectrum with High-Resolution CMB Lensing}
\date{\today}

\author{ \foreignlanguage{vietnamese}{Hồ Nam Nguyễn}}
\affiliation{Physics and Astronomy Department, Stony Brook University, 
Stony Brook, NY 11794}
\author{Neelima Sehgal}
\affiliation{Physics and Astronomy Department, Stony Brook University, 
Stony Brook, NY 11794}
\author{Mathew S. Madhavacheril}
\affiliation{Department of Astrophysical Sciences, Princeton University, 
Princeton, NJ 08544}

\begin{abstract}
We present a method to measure the small-scale matter power spectrum using high-resolution measurements of the gravitational lensing of the Cosmic Microwave Background (CMB). To determine whether small-scale structure today is suppressed on scales below 10 kiloparsecs (corresponding to $M \leq 10^9 M_{\odot}$), one needs to probe CMB-lensing modes out to $L \approx 35,000$, requiring a CMB experiment with about 20 arcsecond resolution or better.  We show that a CMB survey covering 4,000 square degrees of sky, with an instrumental sensitivity of $0.5\mu$K-arcmin at 18 arcsecond resolution, could distinguish between cold dark matter and an alternative, such as 1 keV warm dark matter or $10^{-22}$~eV fuzzy dark matter with about $4\sigma$ significance.  A survey of the same resolution with $0.1\mu$K-arcmin noise could distinguish between cold dark matter and these alternatives at better than $20\sigma$ significance; such high-significance measurements may also allow one to distinguish between a suppression of power due to either baryonic effects or the particle nature of dark matter, since each impacts the shape of the lensing power spectrum differently.  CMB temperature maps yield higher signal-to-noise than polarization maps in this small-scale regime; thus, systematic effects, such as from extragalactic astrophysical foregrounds, need to be carefully considered.  However, these systematic concerns can likely be mitigated with known techniques.  Next-generation CMB lensing may thus provide a robust and powerful method of measuring the small-scale matter power spectrum.
\end{abstract}

\maketitle

\section{Introduction}
\label{sec:intro}
\setcounter{footnote}{0} 

The evidence for the existence of non-baryonic dark matter is compelling~\cite[e.g.,][]{Bullet,planckParams2016}.  A model in which the dark matter consists of a particle that was non-relativistic when structure was growing in the Universe, results in predictions that match observations of structure today on large scales~\cite[e.g.,][]{Peebles1982,Blumenthal1984,Davis1985,Frenk2012,Primack2012,DESCosmology2017}.  We generically call such a model ``cold dark matter'' (CDM).  While the predictions of CDM are well matched to observations on scales of 10 kpc or greater, they are a poor match on scales less than 10 kpc~\cite[e.g.,][]{Ostriker2003,BullockReview}.  Examples of these inconsistencies include: i)~the missing satellites problem, ii)~the too-big-to-fail problem, and the iii)~cusp/core or inner-mass-deficit problem~\cite[e.g.,][]{Brooks2014,Weinberg2015,DelPopolo2017,BullockReview}.  Together these are termed the ``small-scale problems of CDM''.  In the missing satellites problem, the predicted number density of halos is significantly larger than observed for masses below about $10^8 M_\odot$~\cite{Klypin1999,Moore1999}.  The too-big-to-fail problem refers to the observation that high-luminosity satellites comprising the most massive sub-halos of a Milky-Way-size galaxy are much less abundant than predicted by CDM~\cite{Boylan-Kolchin2011}.  For the former problem, one could argue that baryonic physics quenched star-formation, and that those sub-halos actually exist but are dark and currently unobserved.  However, it is much harder to make those arguments for the high-luminosity sub-halos (thus the term "too big to fail").  The cusp/core problem arises because CDM predicts singular density cusps in the centers of halos~\cite{Dubinski1991,Navarro1997}, and observations instead suggest a cored profile in lower-mass systems~\cite{Flores1994,Moore1994}.  While baryonic physics may also resolve this, such a solution is more problematic in systems like dwarf spheroidal galaxies where the baryon fraction is low and dark matter dominates the density~\cite[e.g.,][]{Brooks2014,Oman2015}.   

As a result of these apparent failures of CDM on small scales, a number of alternative dark matter models have been suggested that match CDM predictions on large scales, but that deviate from CDM and instead match observations on small scales.  These models include warm dark matter (WDM)~\cite{Colin2000,Bode2001,Viel2005}, fuzzy dark matter (FDM)~\cite{Turner1983,Press1990,Sin1994,Hu2000,Goodman2000,Peebles2000,Amendola2006,Schive2014,Marsh2016a,Hui2017}, self-interacting dark matter (SIDM)~\cite{Carlson1992,Spergel2000,Vogelsberger2012,Fry2015,Elbert2015,Kaplinghat2016,Kamada2016,TulinYu2017,Huo2017}, and superfluid dark matter (SFDM)~\cite{Berezhiani2015,Khoury2016}, to name a few.  All of these alternative models suppress structure on small scales. However, determining whether one of these models is the correct description of dark matter faces two main challenges:~i)~observations of structure on scales below about 10 kpc are either spotty or open to some interpretation~\cite[e.g.,][]{Oman2016,Hui2017}, and ii)~baryonic physics may also suppress structure on small scales enough to match observations~\cite[e.g.,][]{vanDaalen2011,Brooks2014}, making it difficult to determine which scenario is in play.

To address the challenge of robustly measuring structure on small scales, there are a number of promising avenues being pursued: 

	1.) There are current and planned efforts to search unexplored regions of the Milky Way, to greater depths than achieved before, to find unknown Milky Way satellites.  Current efforts using the Dark Energy Survey (DES) data have yielded a number of new dwarf galaxies~\cite{Koposov2015,Drlica-Wagner2015}, and LSST will find many more.  A challenging aspect of using this method to probe small scale structure, however, is measuring the masses of these dwarf galaxies.  Traditionally, this has required expensive KECK observations to measure the velocity dispersions of the dwarf member stars~\cite{Simon2007}.  Additionally, if a mechanism is quenching star formation in low-mass systems, this method may not provide a complete inventory of small-scale halos.
    
    2.) A similar avenue is to count the number of low-mass galaxies at high redshifts.  Since smaller-mass objects are believed to have formed first, a suppression of small-scale structure is most apparent in the past.  Using galaxy clusters as lenses to magnify background galaxies, the Hubble Frontier Fields team pursued this approach to constrain alternative dark matter models by the abundance of ultra-faint, high-redshift galaxies~\cite{Menci2017}.  This technique has a number of systematic challenges such as the use of photometric redshifts, uncertainties in the selection function of background galaxies, and estimates of the survey volume~\cite{Menci2017}.  A related approach is to measure the number density of high-redshift gamma-ray bursts to constrain alternative dark matter models~\cite{Mesinger2005,deSouza2013}.  This technique has the challenge of determining the mass of the halo hosting the gamma-ray burst~\cite{Mesinger2005,deSouza2013}.  
    
    3.) Another promising technique is to measure substructure in a galaxy by the strong-lensing features it generates, when that galaxy lenses a background galaxy~\cite{Dalal2002}.  This technique is being pursued at optical wavelengths~\cite{Keeton2009,Vegetti2014,Nierenberg2017}, and also at millimeter-wavelengths using ALMA follow-up observations of lensed star-forming galaxies found in CMB surveys~\cite{Hezaveh2013,Hezaveh2016a}.  A challenge to this strong-lensing approach is that large samples are required of such lensed systems before robust constraints can be inferred~\cite{Hezaveh2016b}. Such strong-lensing observations are also expensive. However, LSST and other surveys coming online in the next decade can potentially help in this regard.
    
    4.) Tidal debris streams of stars from disrupted satellites in the Milky Way can also probe the population of sub-halos in the Galaxy~\cite{Moore1999,Johnston2016}.  The gravitational influence of sub-halos can distort and open gaps in these cold stellar streams~\cite{Carlberg2009,Erkal2015}.  However, tidal streams are most influenced by the most massive sub-halos, which can make measuring small-scale structure challenging~\cite{Carlberg2009,Hui2017}.  Currently, the signal-to-noise of detected streams is also low; however, upcoming observations from the Gaia satellite should greatly improve this~\cite{Bovy2014,Mateu2017}.  Lastly, baryonic structures can also distort a stream, making its use as a dark matter probe more complicated~\cite{Amorisco2016}.  
    
    5.) Lyman-$\alpha$ forest observations, which probe the distribution of neutral hydrogen along the line of sight, are also a probe of small-scale structure~\cite[e.g.,][]{Cen1994,Hernquist1996,Croft1999,Hui1999}.  Currently, these observations are in apparent tension with alternative dark matter models that explain locally observed small-scale structure suppression~\cite{Viel2013,Baur2016,Irsic2017}.  However, the Lyman-$\alpha$ probe relies on a baryonic tracer of dark matter, and questions remain regarding whether the baryons themselves can have power on small scales that is not traced by the dark matter~\cite[e.g.,][]{Hui2017}.\\

Given the various challenges of the above-mentioned approaches to measuring small-scale structure, it seems warranted to explore alternative techniques.  In this work, we present a method to measure the small-scale dark matter power spectrum using high-resolution ($\approx 20$ arcseconds) CMB lensing measurements.  One advantage of this technique is that it probes dark matter directly via gravitational lensing, instead of relying on baryonic tracers.  Another advantage is that CMB lensing, on these small scales, is most sensitive to structure at redshifts of 1 to 3.  Since lower mass halos formed first, this makes it more sensitive to suppression of small-scale structure than local probes.  In this work, we show that CMB lensing has the potential to be a powerful and clean tracer of small-scale structure with high statistical significance and minimal systematic uncertainty.  This method may provide a strong complement to the various measurement approaches described above.

To address the issue of distinguishing between a suppression of small-scale structure caused by baryonic physics, as opposed to dark matter alternatives to CDM, we show that the CMB lensing spectrum could potentially have a different shape in the two scenarios.  The high signal-to-noise measurements we forecast here could thus favor one suppression mechanism over the other, if any deviation from the CDM expectation is in fact detected.

In section~\ref{sec:cmblensing}, we briefly summarize the theory of gravitational lensing of the CMB, and in section~\ref{sec:highresexps}, we discuss potential avenues for obtaining the needed high-resolution observations.  In section~\ref{sec:forecasts}, we present statistical forecasts, and in section~\ref{sec:systematics}, we address potential systematic challenges and their mitigation.  We summarize and conclude in section~\ref{sec:discussion}.

\section{CMB Lensing} 
\label{sec:cmblensing}

CMB photons are gravitationally lensed by matter along their path as they traverse the Universe.  The angle by which a CMB photon is deflected is given by the gradient of the projected gravitational potential, ${\bf{d}} = \nabla{\phi}$, where ${\bf{d}}$ is the deflection field.  The projected potential $\phi$ is given by  
\begin{equation}
\phi({\bf{\hat{n}}}) = -2 \int_{0}^{\chi_s} d\chi \frac{D_A(\chi_s-\chi)}{D_A(\chi)D_A(\chi_s)}\Psi(\chi{\bf{\hat{n}}},\chi),
\label{eq:phitheory}
\end{equation}
where $\Psi({\bf{x}},\chi)$ is the three-dimensional gravitational potential~\cite{huokamoto2002}.  
Here $\chi$ is the comoving coordinate distance, $\chi_s$ is the comoving coordinate distance to the last-scattering surface, and $D_A$ is the comoving angular diameter distance.  In a spatially flat universe, $D_A(\chi) = \chi$.  This lensing deflection couples previously uncorrelated Fourier modes of the primordial CMB.  We can thus filter maps of the CMB with an estimator that isolates the specific mode coupling lensing induces to recover an estimate of the projected lensing potential~\cite{blanchard87,bernardeau97,zaldarriaga99,hu2001,huokamoto2002,lewischallinor2006}.   

Taking the power spectrum of the projected potential yields 
\begin{equation}
C_L^{\phi\phi}\approx \frac{8\pi^2}{L^3}\int_0^{\chi_s}\chi d\chi \left(\frac{\chi_s-\chi}{\chi\chi_s}\right)^2P_{\Psi}\left(k,z(\chi)\right)
\end{equation}
for the case of a spatially flat Universe using the  Limber-approximation~\cite{lewischallinor2006}.   
We relate  $P_{\Psi}$ to $P_m$ by 
\begin{eqnarray}
P_{\Psi}&=&\frac{9\Omega_m^2H^4 P_m}{8\pi^2c^4k}\\
&=&\frac{9\Omega_{m0}^2H_0^4 (1+z)^2 P_m}{8\pi^2a^4c^4k}
\end{eqnarray}
where $k$ is the wave number in units of Mpc$^{-1}$, and $P_m$ is the  matter power spectrum in units of Mpc$^3$ \cite{lewischallinor2006,dodelson2003}.
This yields
\begin{equation}
C_L^{\phi\phi}=\frac{9\Omega_{m0}^2H_0^4}{c^4}\int_0^{\chi_s}d\chi\left(\frac{\chi_s-\chi}{\chi^2\chi_s}\right)^2 \frac{(1+z)^2 P_m\left(k,z(\chi)\right)}{k^4}
\end{equation}
where all the units are in comoving coordinates now, and $k\approx \frac{L+0.5}{\chi}$.    
From this we obtain the lensing convergence power spectrum~\cite{hu2001}, $C_L^{\kappa\kappa}$, where
\begin{equation}
C_L^{\kappa\kappa}=\frac{[L(L+1)]^2C_L^{\phi\phi}}{4}.
\end{equation}
$C_L^{\kappa\kappa}$ is the quantity traditionally measured in CMB lensing surveys~\cite[e.g.,][]{actlensing,engelenlensing,plancklensing2014,plancklensing2016}, although it has not yet been measured on the scales needed to probe small-scale structure, which we discuss further in the next section.

\section{High-Resolution CMB Experiments}
\label{sec:highresexps}

\begin{figure}[t]
\centering
\includegraphics[width=0.5\textwidth]{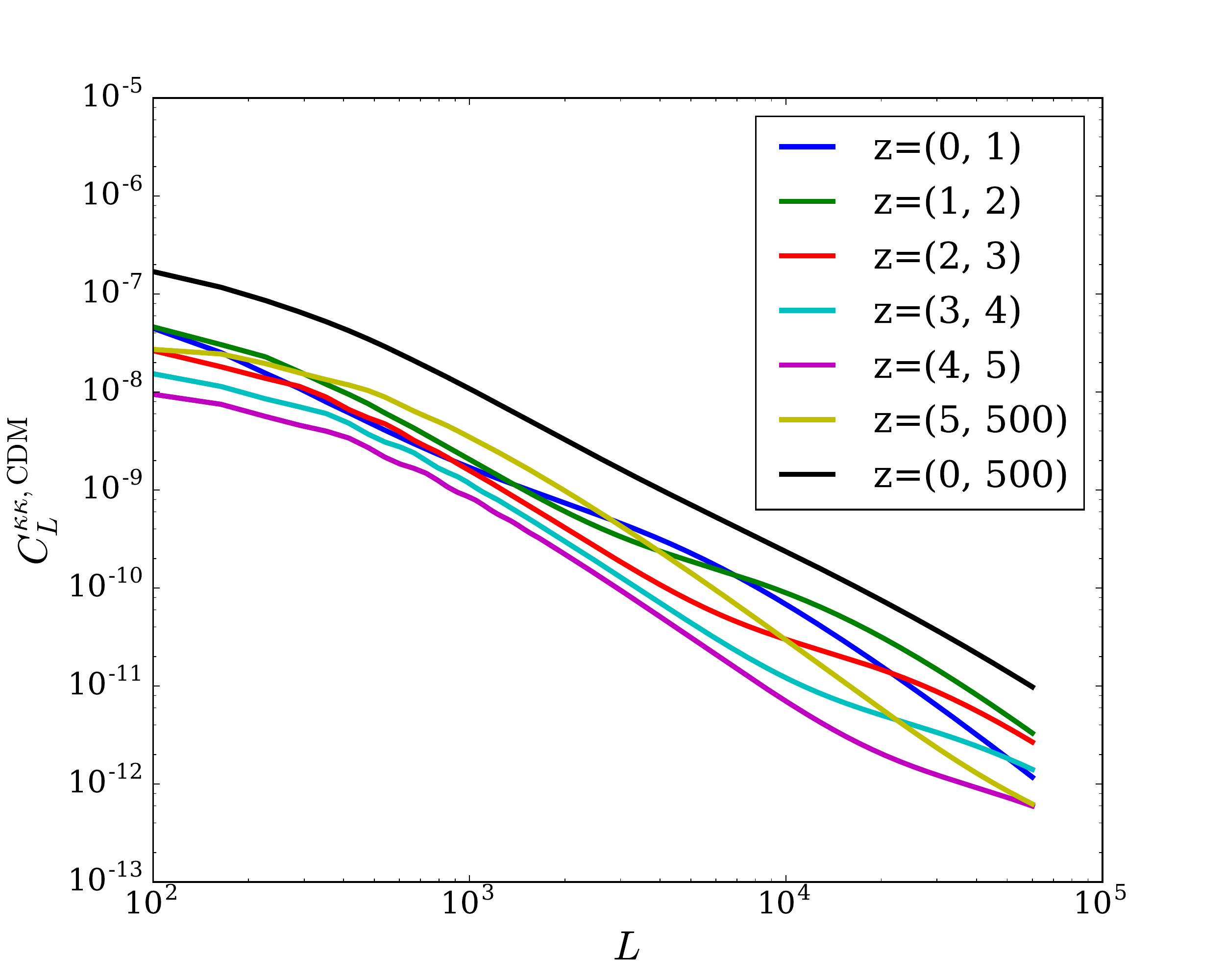}
\caption{Contribution to the lensing power spectra, $C_L^{\kappa\kappa}$, for the CDM model from structure in the given redshift ranges. $C_L^{\kappa\kappa}$ is numerically integrated using the equations in section~\ref{sec:cmblensing} and includes non-linear corrections as discussed in the text.}
\label{fig:power_zrange}
\end{figure}

\begin{figure}[t]
\centering
\includegraphics[width=0.5\textwidth]{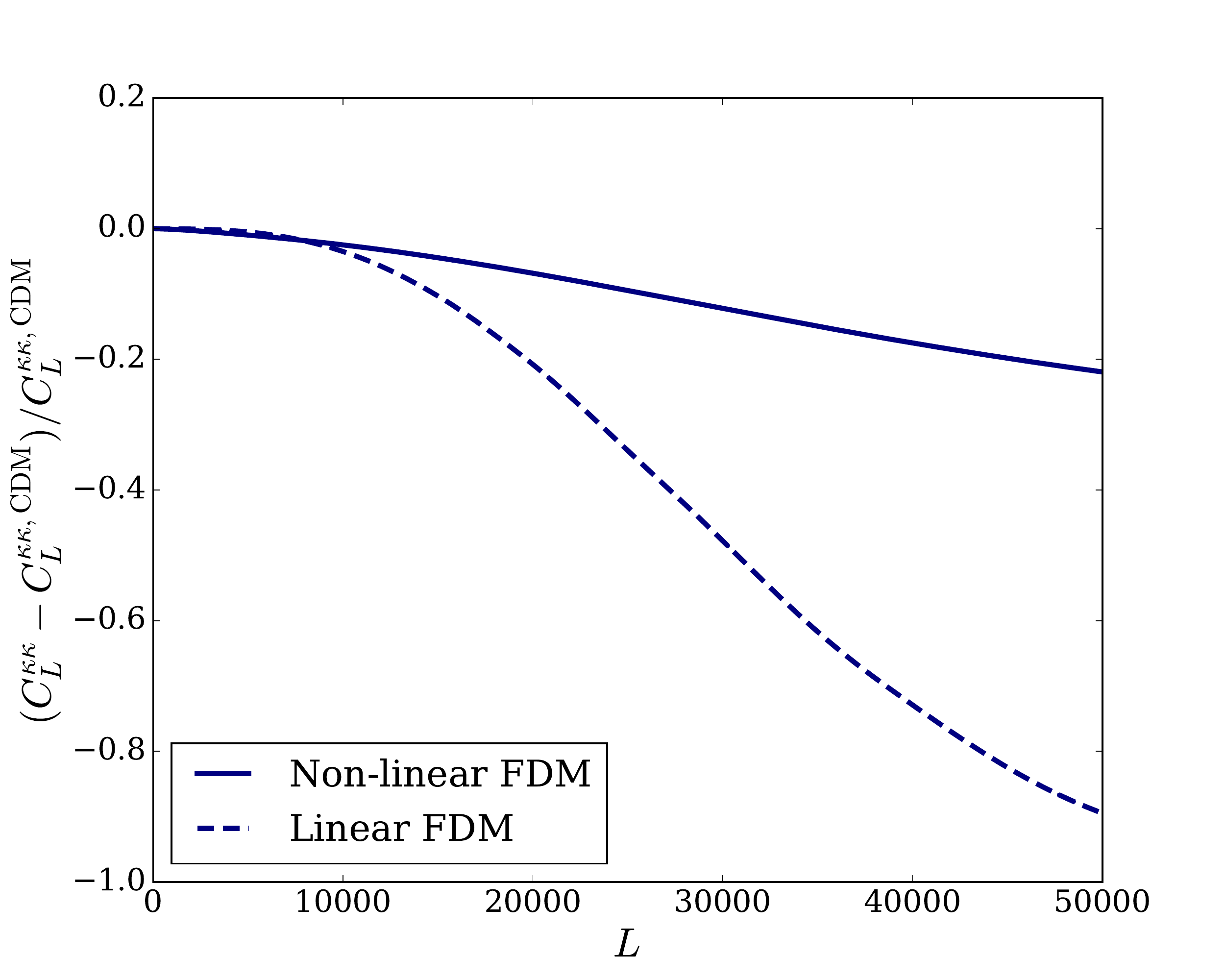}
\caption{Fractional difference in $C_L^{\kappa\kappa}$ between a fuzzy dark matter (FDM) model of mass $m\sim10^{-22}$ eV and the CDM model, with (\textit{blue solid}) and without (\textit{blue dashed}) nonlinear corrections to both models.}
\label{fig:dCkk_nonlinear}
\end{figure}

In order to probe small-scale structure by measuring the matter power spectrum, we need to know which scales at early times collapsed to form structures with $M \leq 10^9$ M$_\odot$, which are on scales less than 10 kpc today.  Considering the epoch of matter-radiation equality, a time when the Universe underwent rapid growth of perturbations, we find that comoving scales of about 150 kpc ($k \approx 10~h$Mpc$^{-1}$) needed to be suppressed then to suppress structures of $M \leq 10^9$ M$_\odot$ today~\cite{Hu2000,Gorbunov2011}.  We note that CMB lensing is most sensitive to structures at $z\approx 2$~\cite{zaldarriaga99}, which is at a comoving distance of $\chi(z=2) \approx 5000$ Mpc. Since $k\approx L / \chi$ as in section~\ref{sec:cmblensing}, to probe comoving scales of $k \approx 7$~Mpc$^{-1}$ with CMB lensing requires measuring lensing $L$-modes of $L \approx 7$~Mpc$^{-1} \times 5000$~Mpc~$= 35,000$.  This is an order of magnitude farther in $L$-modes than CMB lensing surveys have measured to date.  We show in Figure~\ref{fig:power_zrange} the contribution to the lensing power spectrum from structure in different redshift ranges, including non-linear corrections as detailed below.  From this we confirm that for $L \simeq 35,000$, most of the contribution to the lensing power comes from structure between redshifts 1 and 3.

Part of the reason why CMB lensing has not yet probed such small scales is due to the resolution of CMB survey instruments, which have at most of order 1 arcminute resolution~\cite{kosowsky2003,ruhl2004,planckbluebook}.  Using the approximation $\ell_{\rm{max}}\propto\pi/{\rm{(resolution~in~radians)}}$, this gives an $\ell_{\rm{max}} \approx 11,000$. While CMB lensing at any given multipole $L$ is derived from a mix of CMB multipoles $\ell$, at small scales $L$ is derived primarily from multipoles $\ell$ where $\ell \approx L$.  Thus, 1 arcminute resolution translates to $L_{\rm{max}} \approx 11,000$.  To achieve an $L_{\rm{max}} \approx 35,000$, thus requires a resolution of about 20 arcseconds.  Such high resolution, at CMB frequencies, has more traditionally been used by the sub-millimeter community, to study, for example, star-forming galaxies, active galactic nuclei, and proto-planetary systems.  In particular, the proposed Chajnantor Sub/millimeter Survey Telescope (CSST) would consist of a 30-meter dish with 18 arcsecond resolution~\cite{CSST}.  The Large Millimeter Telescope (LMT) in Mexico is already built and operates at 1.1mm to 4mm with a 50-meter dish, achieving 9.5 arcsecond resolution~\cite{LMT}.  As we discuss in the next section, obtaining high-significance measurements of the small-scale matter power spectrum would require putting a CMB camera with the sensitivity of the planned CMB-S4 instrument~\cite{Abazajian2016}, on a dish such as the planned CSST or the existing LMT.

\section{Statistical Forecasts}
\label{sec:forecasts}

\begin{figure}[t]
\centering
\includegraphics[width=0.5\textwidth]{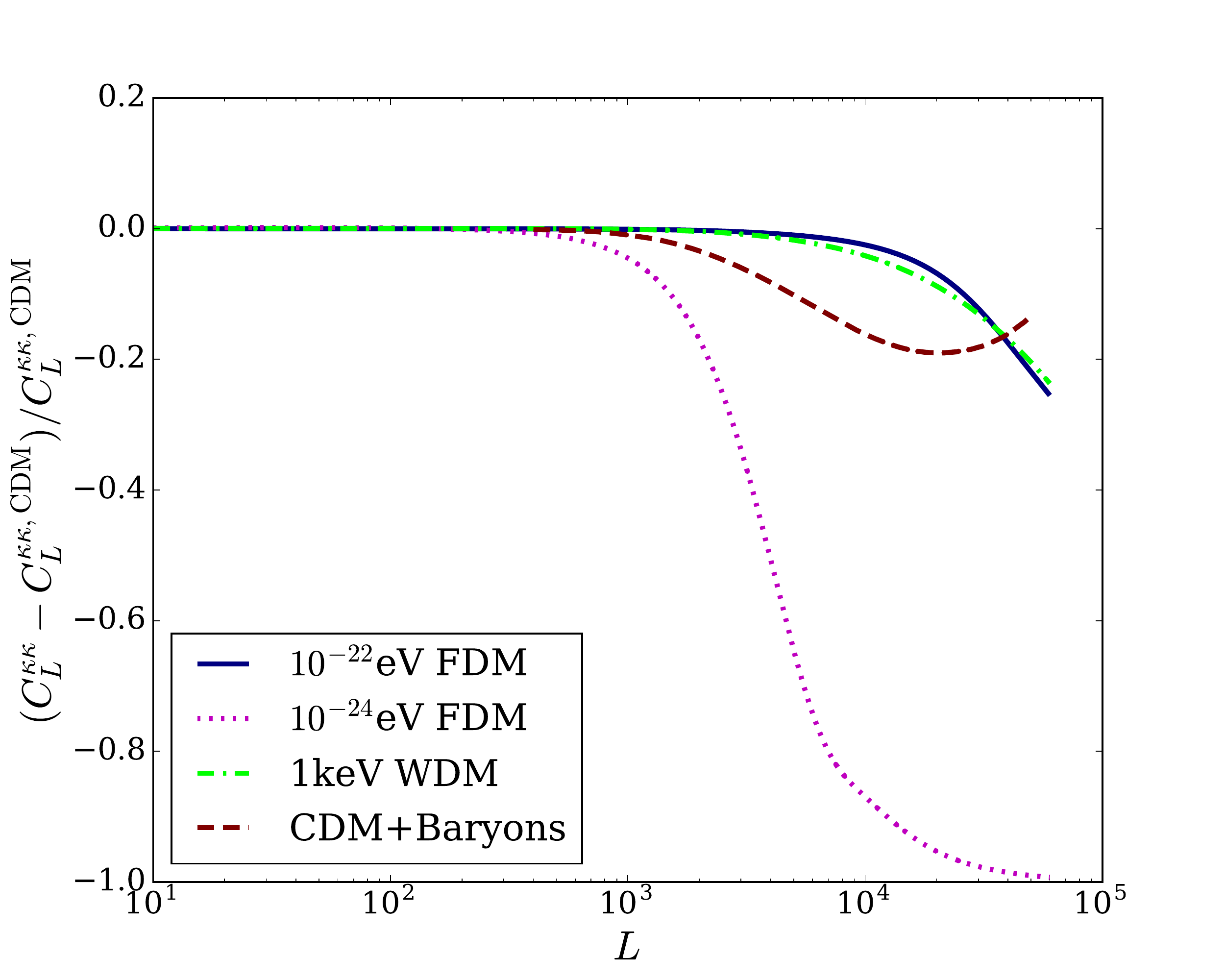}
\caption{Fractional difference in $C_L^{\kappa\kappa}$ between the CDM model and i)~an $m\sim10^{-22}$ eV fuzzy dark matter (FDM) model (\textit{blue solid}), ii)~an $m\sim10^{-24}$ eV FDM model (\textit{magenta dotted}), iii)~a 1keV warm dark matter (WDM) model (\textit{green dash-dotted}), and iv)~a CDM model including baryonic effects (\textit{maroon dashed}).}
\label{fig:dCkk_WF}
\end{figure}

To forecast how well one can distinguish between the CDM prediction of small-scale power and a suppression of small scale power due to alternative dark matter models or baryons, we first calculate the predicted lensing power spectrum for each scenario.  For all calculations below, we assume a fiducial {\it{Planck}}~TT~+~low~P cosmology of $H_0=67.31$ km/s/Mpc, $\Omega_b=0.04904$, $\Omega_m=0.315$, $n_s=0.9655$, and $\sigma_8=0.829$~\cite{planckParams2016}.  We also assume that dark matter consists of either CDM, WDM, or FDM, for example, and not mixtures of these different types.  Figure~\ref{fig:dCkk_nonlinear} shows the fractional difference in CMB lensing power spectra between an FDM and a CDM model of dark matter. The FDM model assumes a dark matter mass of $m=10^{-22}$~eV, which is the mass needed to suppress structure below 1 to 10 kpc~\cite{Hu2000,Hui2017}. Figure~\ref{fig:dCkk_nonlinear} also shows the difference between the linear and non-linear predictions, where the non-linear correction was calculated using the halo model as described in~\cite{Mead2015,meadcode,Hlozek2017}. Here, the lensing power spectra, $C_L^{\kappa\kappa}$, for FDM and CDM are obtained by using the Limber approximation and the equations in section~\ref{sec:cmblensing} to integrate the matter power spectrum obtained from the~\textit{WarmAndFuzzy} code~\cite{Marsh2016}. The $C_L^{\kappa\kappa}$ for Figure~\ref{fig:power_zrange} are obtained in the same way. For the analysis presented in this work, we use power spectra integrated over the redshift range from 0 to 500, shown by the black curve in Figure~\ref{fig:power_zrange} for the CDM case.  

In Figure~\ref{fig:dCkk_WF}, we show the fractional difference in lensing power for the $10^{-22}$~eV FDM model compared to CDM, and for a 1~keV WDM model compared to CDM. The lensing power spectrum for the WDM model is also obtained using the~\textit{WarmAndFuzzy} code.  The shapes of the lensing power spectra for FDM and WDM are very similar; thus, we will show results for the $10^{-22}$~eV FDM model and view them as applicable to a 1~keV WDM model as well. We also show the fractional difference for a $10^{-24}$~eV FDM model, for reference.  Since baryonic effects can also suppress small-scale structure, we show the prediction for one CDM+baryons model using the published $P_m(k)$ from~\cite{vanDaalen2011}, obtained from the OWLS simulations~\cite{Schaye2010}.  While this represents just one model of baryonic effects, we note that the shape of $\Delta C_L^{\kappa\kappa}$ is markedly different from the fiducial FDM/WDM model.  The difference in the shapes of the lensing power spectra may provide a promising way of determining the mechanism of small-scale structure suppression, if any deviation from the CDM prediction is found. We emphasize, however, that our primary motive in this work is to devise a way to make a robust measurement of the small-scale matter power spectrum, and to see whether or not that matches the CDM prediction. That will necessarily inform baryonic and dark matter physics, potentially in ways we cannot predict.

\begin{figure}[t]
\centering
\includegraphics[width=0.5\textwidth]{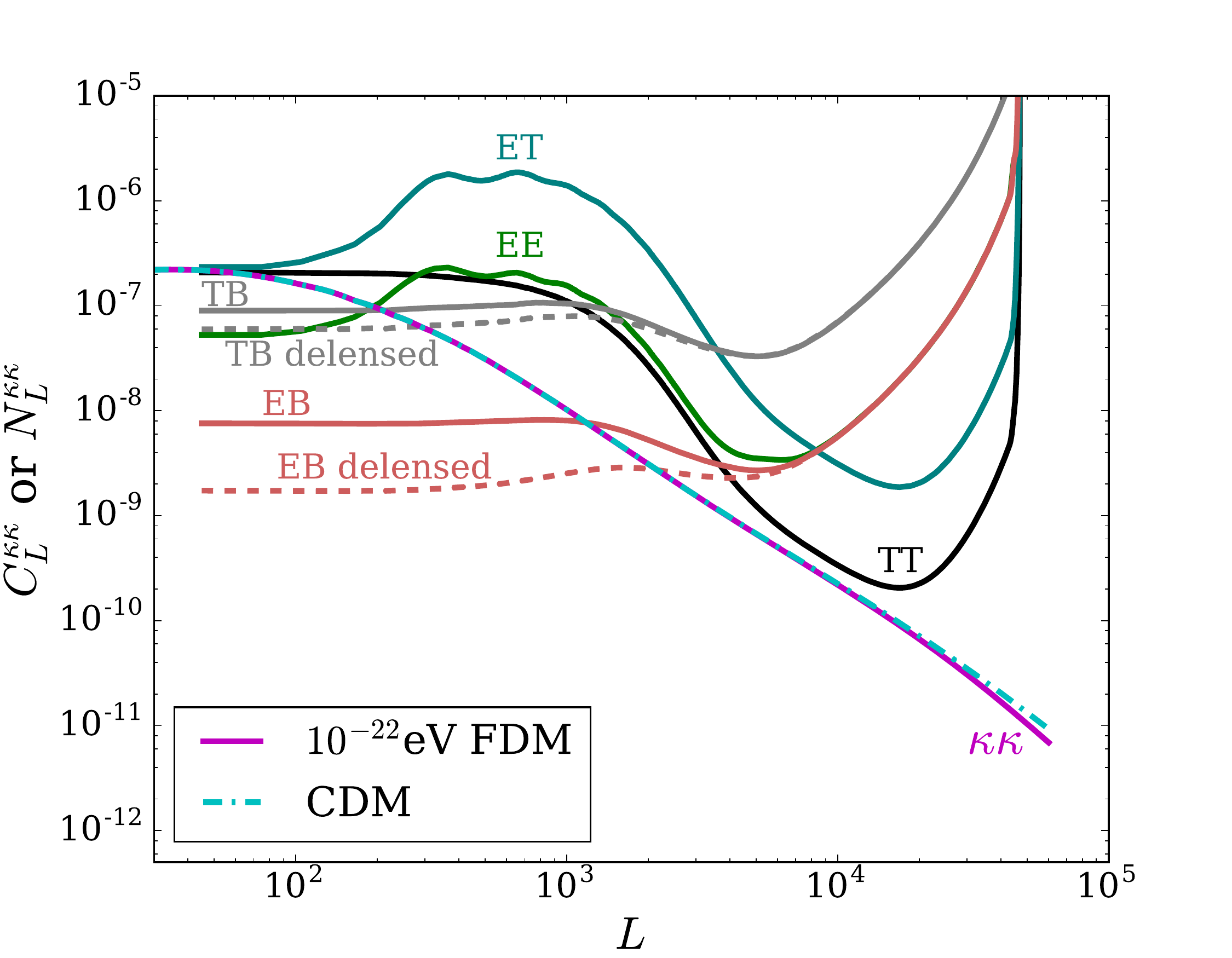}
\caption{Lensing convergence power, $C_L^{\kappa\kappa}$, compared to noise power, $N_L^{\kappa\kappa}$, following~\cite{hudedeovale2007} for different map combinations used in the quadratic estimator for lensing reconstruction. All the noise spectra correspond to an experiment with an 18'' beam and 0.1 $\mu$K-arcmin noise in temperature.  For $L$'s above $10^4$, where most of the signal-to-noise resides when measuring a deviation from CDM on small scales, the $TT$ estimator has lower noise than $EB$. Figure~\ref{fig:covMatDiagonals} in the appendix shows $N_L^{\kappa\kappa}$ for $TT$ from simulations using the same quadratic estimator.  Simulations pick up excess noise at $L \simeq 10^4$ not modelled in~\cite{hudedeovale2007}, as detailed in the appendix.} 
\label{fig:ClkkNlkk}
\end{figure}

For forecasting, we employ simulations, which we describe in detail in the appendix, to capture the details of the noise covariance matrix.  However, to gain qualitative insight, we also calculate the noise power spectra, $N_L^{\kappa\kappa}$, following~\cite{hudedeovale2007}, which we show in Figure~\ref{fig:ClkkNlkk}.  Here, the $N_L^{\kappa\kappa}$ are derived assuming a standard quadratic estimator to estimate $C_L^{\kappa\kappa}$, following~\cite{hudedeovale2007}.  This estimator differs from the estimator of~\cite{huokamoto2002} in that it is tailored to measure smaller halo lenses than the original estimator. This is because it takes advantage of the fact that the lensing signal of these smaller halos appears as a perturbation on top of a smooth CMB background gradient~\cite{Dodelson2003b,Holder2004,LewisKing2006}. 

Specifically, we estimate the lensing convergence field, $\kappa$, by making two versions of filtered data.  One version is filtered to isolate the CMB background gradient, and the other version is filtered to isolate the CMB fluctuations on small scales.  The former is constructed by taking the weighted gradient of the lensed CMB map
\begin{equation}
{\bf G}_{\Bell}^{TT} = i \, {\Bell}\, W_l^{TT} \,T_{\Bell},
\end{equation}
where the weight filter is
\begin{equation}
W_l^{TT} = \tilde{C}_l^{TT} (C_l^{TT}+N_l^{TT})^{-1}.
\end{equation}
for $l \leq l_{\rm{G}}$, and $W_{l}^{TT} = 0$ for $l > l_{\rm{G}}$.
Here we use temperature maps as an example, and note that $ \tilde{C}_l$ and ${C}_l$ are the unlensed and lensed CMB power spectra respectively from a fiducial theoretical model.  $N_l$ is the noise power spectrum. The $l_{\rm{G}}$ is a cutoff scale and is set to $l_{\rm{G}}=2000$ to eliminate biased potential reconstructions of massive halos contributing to the lensing power spectrum, as was done in~\cite[e.g.,][]{Madhavacheril2015}. This ``gradient leg'' cutoff scale comes at the cost of some signal-to-noise in the power spectrum, but only on large scales that are not relevant to this work.  

The second filtered map is an inverse-variance weighted map given by
\begin{equation}
L^T_{\Bell} = W_l^T\,T_{\Bell},
\end{equation}
where
\begin{equation}
W_l^T = (C_l^{TT}+N_l^{TT})^{-1}.
\end{equation}
We then take the divergence of the product of these two maps, following~\cite{hudedeovale2007}, to obtain an estimate of $\kappa$ as follows,

\begin{equation}
\frac{\kappa_{L}^{TT}}{A_L^{TT}} = -\int {\rm d}^2 \hat{\bf n}\,e^{-i \hat{\bf n} \cdot {\Bell}} \, \left\{ \nabla \cdot [{\bf G}^{TT}(\hat{\bf n}) \,L^{T}(\hat{\bf n})]\right\}.
\label{eq:main}
\end{equation}
Here the real-space lensing convergence field constructed from temperature data is
\begin{equation}
\kappa^{TT}(\hat{\bf n}) = \int \frac{{\rm d}^2 L}{(2\pi)^2}\,e^{i {L} \cdot \hat{\bf n}}\,\kappa_{L}^{TT}.
\end{equation}
The normalization factor is given by
\begin{equation}
\frac{1}{A_L^{TT}} = \frac{2}{L^2}\, \int \frac{{\rm d}^2 l_1}{(2\pi)^2}\, [{L} \cdot {\Bell}_1] \, W_{l_1}^{TT}\,W_{l_2}^T\, f^{TT} ({\Bell}_1, {\Bell}_2),
\label{eq:norm}
\end{equation}
with
\begin{equation}
f^{TT} ({\Bell}_1, {\Bell}_2) = [{L} \cdot {\Bell}_1]\tilde{C}^{TT}_{l_1} + [{L} \cdot {\Bell}_2]\tilde{C}^{TT}_{l_2}
\end{equation}
and ${L}={\Bell_1} + {\Bell_2}$.

In Figure~\ref{fig:ClkkNlkk}, we show $N_L^{\kappa\kappa}$ using five different CMB map combinations ($TT, EE, ET, TB$, and $EB$), where $T, E$, and $B$ represent temperature, E-mode, and B-mode CMB maps, respectively.  Each of these $N_L^{\kappa\kappa}$ curves shows the noise per mode, assuming 18'' resolution (to match the planned CSST), 0.1~$\mu$K-arcmin instrumental white noise in temperature, and 0.1~$\times~\sqrt{2}\mu$K-arcmin white noise in polarization.  The $N_L^{\kappa\kappa}$ are derived following \cite{hudedeovale2007}, where 
\begin{equation}
N_L^{\kappa\kappa,XY}=\frac{L^2}{4}N_L^{dd,XY},
\label{eq:NLkk}
\end{equation}
and
\begin{equation}
(N_L^{dd,XY})^{-1} =\frac{2}{L^2}\int\frac{d^2\ell_1}{(2\pi)^2}\dotfac{1}W^{XY}_{\ell_1}W^Y_{\ell_2}c^Yf^{XY}
\label{eq:NL}
\end{equation}
for $X,Y\in T,E,B$.  The latter equation is an integral over all CMB modes, $\Bell_1$, with the constraint that $\Bell_2 = \bL - \Bell_1$. The terms $W^{XY}_{\ell_1}, W^Y_{\ell_2}, c^Y$, and $f^{XY}$ are defined in \cite{hudedeovale2007}.  We note that for the small scales investigated in this work, it is likely possible to construct a more optimal maximum likelihood estimator~\cite{Raghunathan2017,FerraroSherwin2017,Horowitz2017}. Here, we use the quadratic estimator described above, and treat our forecasts as potentially conservative.  

From Figure~\ref{fig:ClkkNlkk}, we see that for measuring $C_L^{\kappa\kappa}$ on scales below $L\approx 2000$, the $EB$ estimator from polarization maps has the lowest noise.  This noise can be further reduced by iteratively delensing the B-mode map, as shown by the dashed curves~\cite{seljakhiratadelens2004,smithdelens2012}.  However, for probing scales of order $L\approx 10,000$, the $TT$ estimator is better.  The reason the $TT$ noise decreases so significantly at small scales is because at these scales the power in the lensing signal dominates over the power in the primordial temperature anisotropy.  A similar effect happens for the polarization maps, however at 0.1~$\times \sqrt{2}\mu$K-arcmin noise levels, the CMB signal does not dominate over instrument noise at these scales.  As a result, performing the lensing potential reconstruction using temperature maps yields the largest signal-to-noise ratio (SNR).  In the forecasts that follow, we assume only temperature maps are used in the lensing reconstruction since including the other estimators only marginally improves the results.  

\begin{figure}[t]
\centering
\vspace{-6mm}
\includegraphics[width=0.5\textwidth]{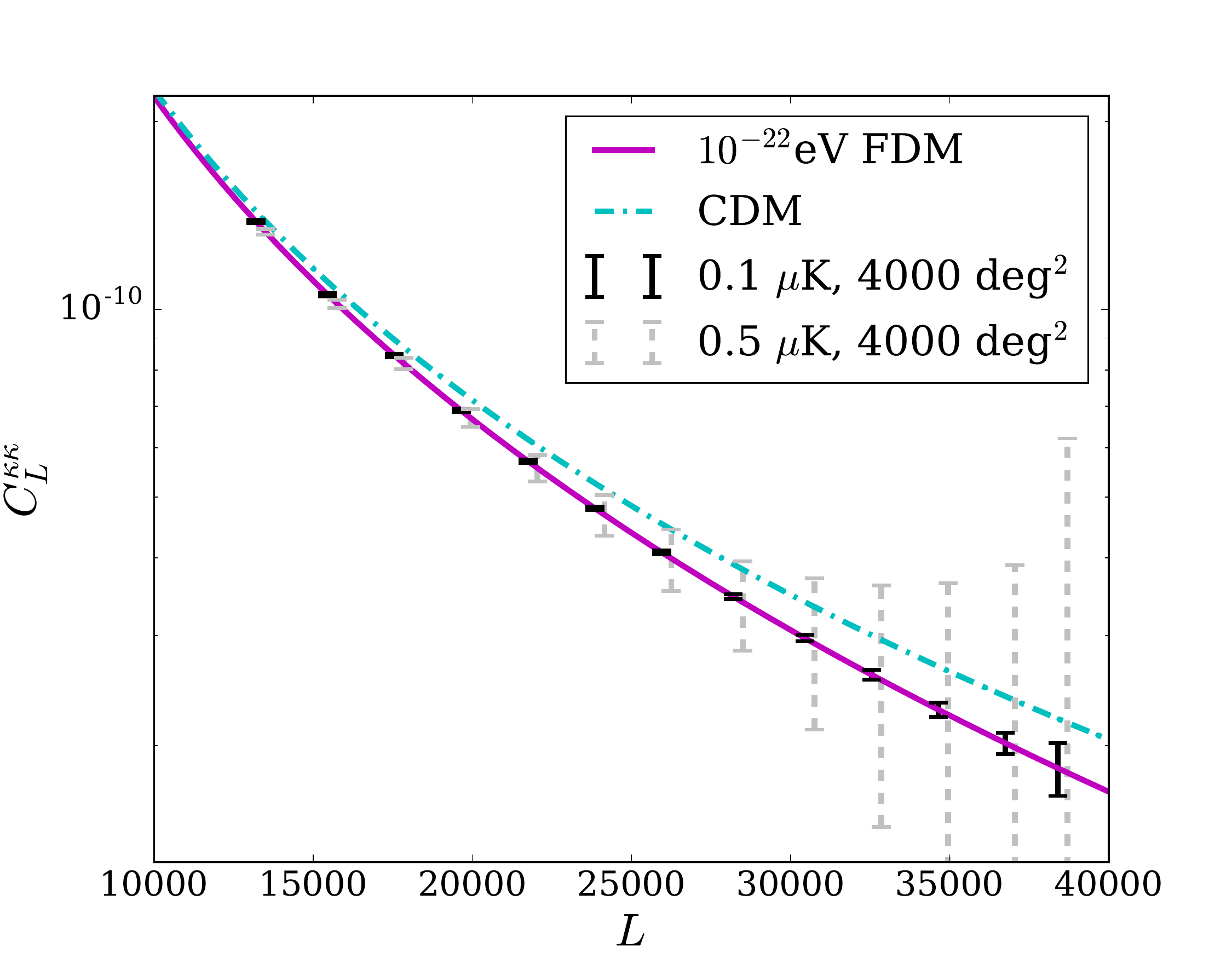}
\caption{Lensing convergence power spectra of an $m\sim 10^{-22}$~eV FDM model and a CDM model. The error bars shown are from the diagonal terms of the simulation-based \T\T\ lensing noise covariance matrix described in the appendix.  Here the \textit{black solid} and \textit{silver dashed} error bars correspond to 0.1 $\mu$K-arcmin  and 0.5 $\mu$K-arcmin CMB noise in temperature, respectively. For both sets of error bars, a 10\% observed sky fraction is assumed and 18'' resolution.  Note that the 0.5 $\mu$K-arcmin error bars are shifted to the right for clarity.}
\label{fig:errorbar}  
\end{figure}

We calculate the SNR with which we could distinguish between CDM and an alternative model for the lensing power spectrum, such as FDM, as
\begin{equation}
\frac{S}{N}=\sqrt[]{\sum_{L,L'}(X_L-Y_L){\rm C}^{-1}_{LL'}(X_{L'}-Y_{L'})}
\end{equation}
where $X_L = C_L^{\kappa\kappa,\rm FDM}$, $Y_L= C_L^{\kappa\kappa,\rm CDM}$, and ${\rm C}^{-1}_{LL'}$ is an element of the inverted covariance matrix corresponding to row $L$ and column $L'$. 
For the $N_L^{\kappa\kappa}$ from the quadratic estimator described in~\cite{hudedeovale2007}, on large lensing scales $L$ traditionally measured, treating each $L$-mode as independent is a good approximation~\cite{Hanson2011}. However, each $L$-mode is not independent on the small scales considered here.  This is because the primordial background CMB gradient enters as a source of sample variance noise.  It may be possible for maximum likelihood estimators under development to utilize knowledge of the background CMB gradient, and remove it as a source of noise in the estimator~\cite{Raghunathan2017,FerraroSherwin2017,Horowitz2017}. However, in this work, we adopt the quadratic estimator in~\cite{hudedeovale2007} and construct the full noise covariance matrix, including off-diagonal terms, using simulations.  We describe the simulations and the construction of the covariance matrix in detail in the appendix.

\begin{center}
\renewcommand{\arraystretch}{0.7} 
\begin{table}[t]
\begin{tabularx}{\linewidth}{C|C|C|C}
	\hline
	\hline
 	Sky fraction & Noise & \multicolumn{2}{c}{Signal-to-noise ratio}   \\
    (f\textsubscript{sky})& ($\mu$K-arcmin) & 18'' Resolution & 9.5 '' Resolution\\
    \hline    
    0.1 & 0.5 & 3.9 & 5.2 \\ 
	0.025 & 0.1 & 10.1 & 15.9 \\
	0.1 & 0.1 & 20.2 & 31.9 \\
	\hline 
\end{tabularx}
\caption{Significance with which an $m\sim10^{-22}$~eV FDM model can be distinguished from a CDM model, based on observations of high-resolution CMB lensing.  Here we vary observed sky fraction, noise levels in temperature, and resolution. The lensing noise power assumes only the \T\T\ estimator is used, however, the gain from including other estimators is minimal. For these signal-to-noise ratios, we use the full simulation-based lensing noise covariance matrix detailed in the appendix.}
\label{tab:sn}
\end{table}
\end{center}

In Figure~\ref{fig:errorbar}, we show as error bars on $C_L^{\kappa\kappa}$ the diagonal terms of the simulation-based noise covariance matrix for $TT$. Here, we assume a survey of $10\%$ of the sky (4,000 square degrees), at 18'' resolution, with $0.5\mu$K-arcmin (grey), and $0.1\mu$K-arcmin (black) white noise levels.  Table~\ref{tab:sn} shows the SNRs for these two cases, as well as for a survey covering less than $3\%$ of the sky (1,000 square degrees). We limit the CMB-$\ell$ range from 100 to 45,000 since the inclusion of more modes does not make any significant impact on the SNRs. From this we see that a survey covering 4,000 square degrees of sky at a noise level of $0.5\mu$K-arcmin can already detect the difference between $10^{-22}$~eV FDM and CDM with almost $4\sigma$ significance.  For deeper noise levels of $0.1\mu$K-arcmin, SNRs over 20 can be achieved. With finer resolution, such as 9.5'' to match the LMT, SNRs above 30 are possible.   

\begin{figure}[t]
\centering
\includegraphics[width=0.5\textwidth]{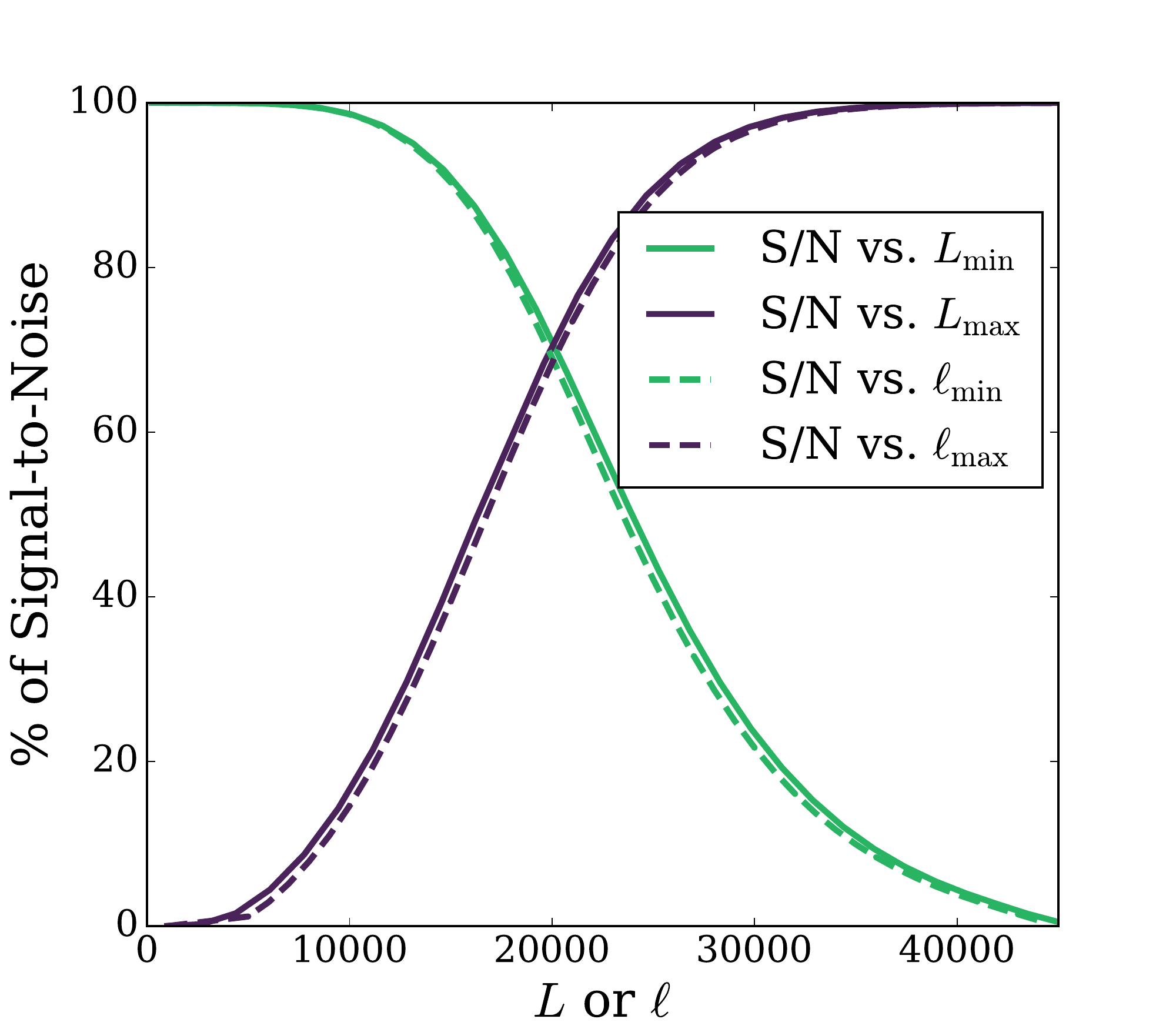}
\caption{Percentage of total signal-to-noise ratio in distinguishing an $m\sim10^{-22}$ eV FDM model from a CDM model, as a function of maximum and minimum CMB and CMB lensing multipole moments ($\ell$ and $L$ respectively). The lower bounds are fixed to 100 when the upper bounds are varied, and the upper bounds are fixed to 45,000 when the lower bounds are varied. This is shown for the fiducial case of 0.1 $\mu$K-arcmin CMB noise in temperature, as in Figure~\ref{fig:errorbar}.  Here, we calculate the signal-to-noise ratio using the $N_L^{\kappa\kappa}$ from Eq.~\ref{eq:NL} and assume independent $L$-modes, instead of using the full covariance matrix discussed in the text, to gain qualitative insight.}  
\label{fig:snAll}
\end{figure}

\begin{figure}[t]
\centering
\includegraphics[width=0.5\textwidth]{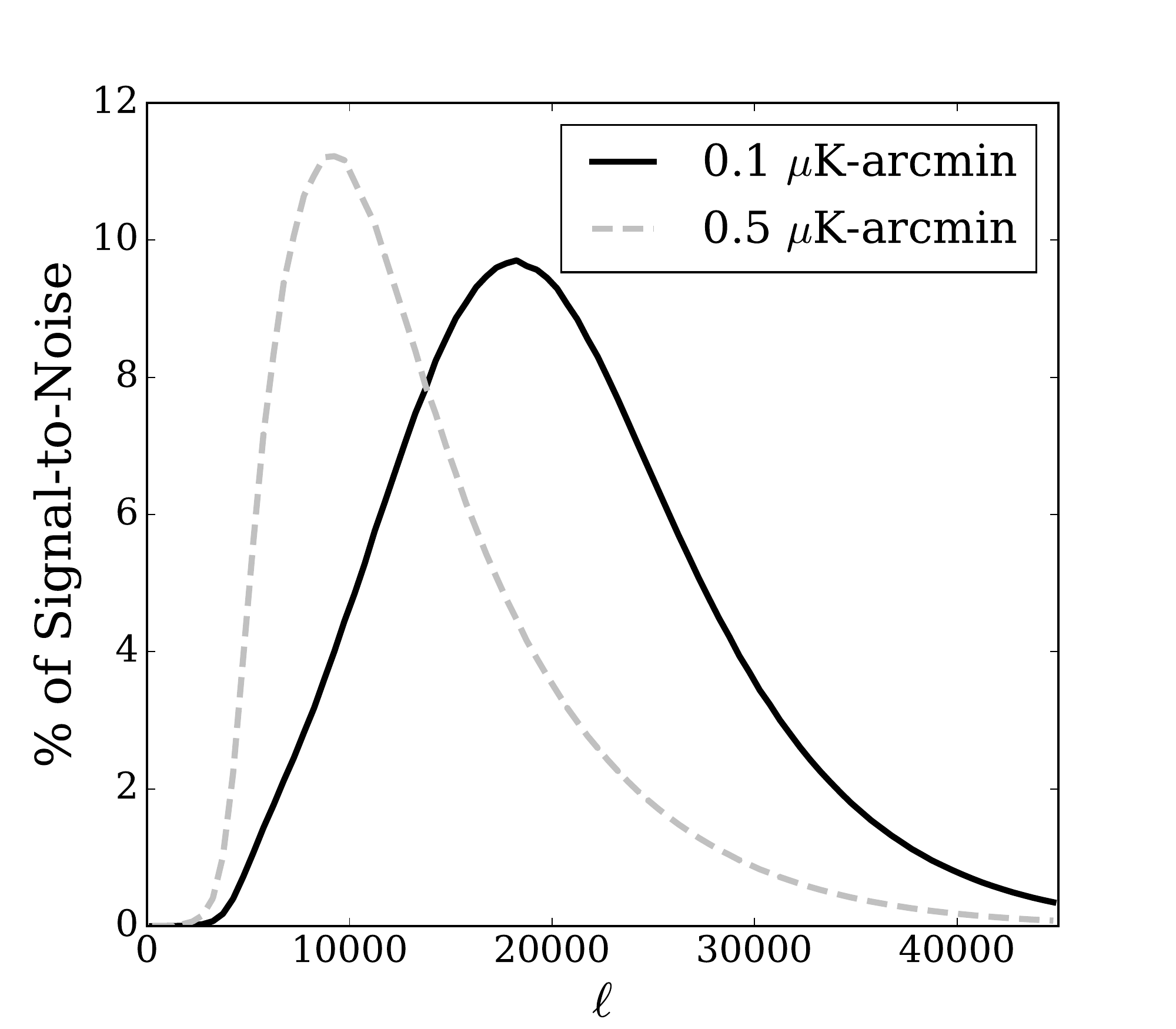}
\caption{Fractional contribution of signal-to-noise ratio as a function of CMB multipole moment $\ell$.  As in Figure~\ref{fig:snAll}, 18'' resolution and $N_L^{\kappa\kappa}$ from Eq.~\ref{eq:NL} are used to gain qualitative insight. For higher noise levels, more of the weight comes from lower CMB $\ell$'s.  See text for details.}
\label{fig:snBins}
\end{figure}

To see which lensing $L$-modes and CMB $\ell$-modes contribute most to the SNR, we show in Figure~\ref{fig:snAll}, for lensing $L$-modes (solid) or CMB $\ell$-modes (dashed), the SNR as a function of minimum and maximum modes included in the calculation.  Here the maximum $\ell$-mode refers to the maximum multipole used in the CMB map that was filtered to isolate the small-scale CMB fluctuations, as discussed above.  In this Figure and in Figure~\ref{fig:snBins}, we use the $N_L^{\kappa\kappa}$ from Eq.~\ref{eq:NL} and assume no off-diagonal terms in the covariance matrix, to gain qualitative insight.  Using a full simulation-based covariance matrix gives a similar result, but is more computationally expensive when exploring many $\ell$-mode ranges. In Figure~\ref{fig:snAll}, the lower bounds are fixed to $\ell/L=100$, when the upper bounds are varied, and the upper bounds are fixed to $\ell/L=45,000$ when the lower bounds are varied.  This is shown for the fiducial case of $0.1\mu$K-arcmin noise and 18'' resolution.  The SNR stops increasing at around $\ell/L=30,000$, consistent with the rise in the noise curves shown in Figure~\ref{fig:ClkkNlkk}.  The SNR only starts increasing significantly when $\ell/L = 10,000$, which is the multipole where the $10^{-22}$~eV FDM $C_L^{\kappa\kappa}$ makes a notable deviation from that of CDM, as seen in Figure~\ref{fig:errorbar}.  To further identify which $\ell$-modes contribute to the SNR, we divide the $\ell$-range into bins of width $\Delta \ell=500$.  For each bin, we estimate the noise curve $N_L^{\kappa\kappa}$ using just the $\ell$-modes from that bin, and calculate the SNR.  This is shown in Figure~\ref{fig:snBins}, for the $0.1\mu$K-arcmin (solid) and $0.5\mu$K-arcmin (dashed) cases shown in Figure~\ref{fig:errorbar}.  We note that this process of dividing into $\ell$ bins recovers 49$\%$ and 46$\%$ of the total SNR for the 0.1 and $0.5\mu$K-arcmin cases, respectively.  The reason it is not 100$\%$ is because we are not allowing $\bL$ modes derived from $\Bell_1$ and $\Bell_2$ pairs spanning two $\ell$ bins.  We see that as the noise level decreases, more of the SNR comes from higher $\ell$-modes.  We also find from Figures~\ref{fig:snAll} and~\ref{fig:snBins} that most of the SNR comes from $\ell$-modes where $\ell \in (10,000,~30,000)$, with a peak at $\ell \approx 20,000$ for $0.1\mu$K-arcmin noise.  For $0.5\mu$K-arcmin noise, the SNR is mostly from $\ell \in (5,000,~25,000)$ and peaks at $\ell \approx 9,000$.\\

\section{Tests of Parameter Degeneracies}
\label{sec:degeneracies}
The SNRs shown in Table~\ref{tab:sn} assume fixed cosmological parameters given by the fiducial cosmology in section~\ref{sec:forecasts}, and are calculated by only varying the dark matter model away from CDM.  Given that the SNRs of neutrino mass measurements, made by exploiting a similar suppression of the CMB lensing power spectrum, have a known limit due to the uncertainty on the optical depth parameter $\tau$~\cite[e.g.,][]{Abazajian2016}, we explicitly check whether such a limit applies to this case as well.  We perform a Fisher Matrix analysis varying $A_s$, $n_s$, $\tau$, $\Omega_bh^2$, $\Omega_{\rm FDM}h^2$, $H_0$, and $m_{\rm FDM}$.  We set the fiducial values of these parameters to be $H_0=67.31$ km/s/Mpc, $\Omega_bh^2=0.0222$, $\Omega_{\rm FDM}h^2=0.1197$, $n_s=0.9655$, $A_s=2.2\times10^{-9}$, $\tau=0.06$ and $m_{\rm FDM}=10^{-22}$~eV.  To calculate the derivatives for the Fisher Matrix, we use step sizes that are $1\%$ of the corresponding fiducial value.  ... All power spectra are obtained via AxionCAMB~\cite{Hlozek2015}, a modified version of CAMB~\cite{Lewis2000} that incorporates FDM.  We use AxionCAMB because it allows $\tau$ and $A_s$ to be explicitly varied, unlike the~\textit{WarmAndFuzzy} code~\cite{Marsh2016}, however only the latter include non-linear corrections.  Thus, we do the Fisher analysis with linear spectra, and expect that the qualitative behavior will be the same in the non-linear case.  We expect using the linear power spectrum is suitable for exploring these degeneracies because $A_s$, $\tau$, and neutrino mass effect much larger scales, which are largely linear, than $m_{\rm FDM}$; thus the degeneracy behavior we are exploring will not change if we use a non-linear matter power spectrum on small scales where these three parameters have negligible impact.  We also use only the diagonal terms of $N_L^{\kappa\kappa}$ from Eqs.~\ref{eq:NLkk} and~\ref{eq:NL}, instead of the full simulation-based covariance matrix, for ease of calculation, and expect the qualitative behavior to be unchanged.  

In the Fisher analysis, we assume the high-resolution experiment to survey $10\%$ of the sky with 0.1~$\mu$K-arcmin noise and 18" resolution, and include CMB $\ell$-modes from $\ell=100$ to 45,000.  We also include data from the planned CMB-S4 survey covering $40\%$ of the sky with 1.0~$\mu$K-arcmin noise and 2' resolution, measuring $\ell=30$ to 5000~\cite{Abazajian2016}.  In addition, we include \planck 2015 priors~\cite{planckParams2016}, except for the error on $\tau$, which we allow to vary.  We find that $\sigma(m_{\rm FDM})$ is insensitive to decreasing the uncertainty on $\tau$ from $\sigma(\tau)=0.01$ (the current uncertainty) to $\sigma(\tau)=0.002$ (the cosmic variance limit).  We can understand this intuitively because non-zero neutrino mass suppresses $C_L^{\kappa\kappa}$ at the lowest $L$-modes, whereas FDM with a mass of $m_{\rm FDM}=10^{-22}$~eV only starts suppressing $C_L^{\kappa\kappa}$ at $L>1000$, as seen in Figure~\ref{fig:dCkk_WF}.  The lack of suppression for $L<1000$ gives a long lever arm between $L\simeq 100$ to 1000 within which to measure the unsuppressed power.  This makes these dark matter measurements insensitive to the prior on $\tau$, in contrast to neutrino mass measurements.

\begin{figure}[t]
\centering
\includegraphics[width=0.5\textwidth]{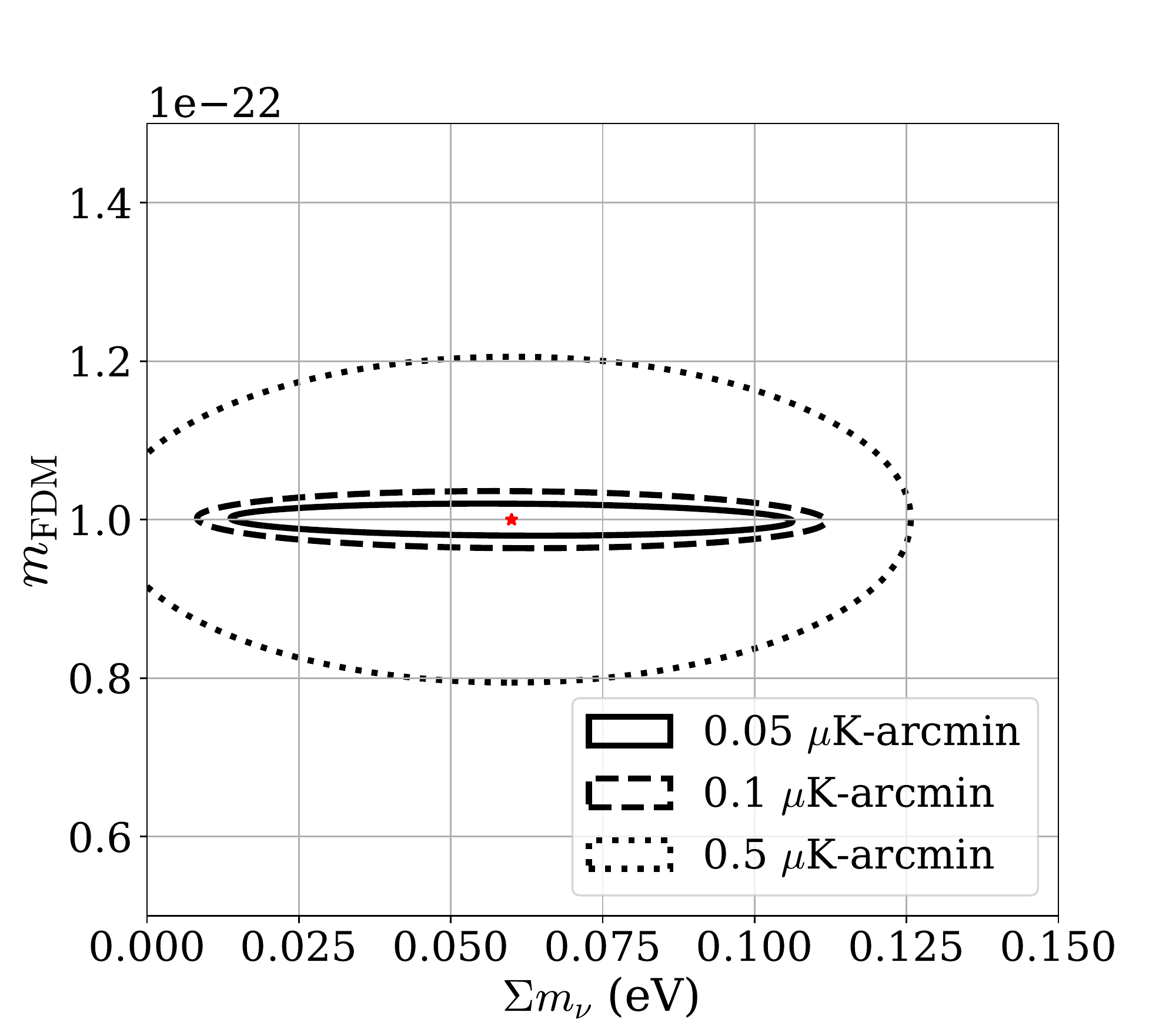}
\caption{Joint constraint (68\% CL) on $m_{\rm FDM}$ and $\Sigma m_{\nu}$ for different noise levels of a high-resolution experiment with 18'' resolution over 10\% of sky. Here we include \planck 2015 priors for the 6 base $\Lambda$CDM parameters and assume CMB-S4 data will be available (1.0~$\mu$K-arcmin noise in temperature and 2' resolution over 40\% of sky). BAO is not included in the Fisher Matrix analysis. We note there is minimal degeneracy between $m_{\rm FDM}$ and $\Sigma m_{\nu}$.}
\label{fig:confEllipse}
\vspace{-3mm}
\end{figure}

We also explicitly check the degeneracy between $m_{\rm FDM}$ and the sum of neutrino masses, $\Sigma m_{\nu}$. We perform a similar Fisher Matrix analysis as above, now varying the additional parameter $\Sigma m_{\nu}$ and assuming the \planck 2015 prior on $\tau$~\cite{planckParams2016}. The result is shown in Figure~\ref{fig:confEllipse} via 68\% confidence level ellipses for the joint constraint on $m_{\rm FDM}$ and $\Sigma m_{\nu}$, for different noise levels for the high-resolution experiment. We observe that the ellipses remain practically horizontal as the noise level decreases, indicating minimal degeneracy between $m_{\rm FDM}$ and $\Sigma m_{\nu}$.

\section{Systematic Considerations}
\label{sec:systematics}

There are a number of potential systematic effects that need to be considered when measuring the lensing power spectrum in this uncharted, small-scale regime.  These systematic effects include biases from Galactic and extragalactic foregrounds, mis-subtraction of the Gaussian bias term of the lensing power spectrum, and additional mode-coupling signals that may contaminate the lensing signal when we measure the CMB with instrumental noise levels as low as about $0.1\mu$K-arcmin. \\ 

{\it{Gaussian Noise Bias:}} Gaussian noise arising from the primary CMB is present when one reconstructs the lensing potential. Calculating the power spectrum of that reconstruction, thus results in a large bias to the lensing power spectrum, usually called $N_0$ bias~\cite{huokamoto2002}. A key to accurately subtracting off this bias is employing simulations that match the data to within about $10\%$ in power~\cite{Namikawa2013, Sherwin2017}.  For measuring the lensing power spectrum on scales an order of magnitude smaller than achieved to date ($\ell \sim 30,000$), it is not realistic to assume that any simulations will match the data to that level of accuracy.  The reason is that the small-scale matter power spectrum can vary by more than $10\%$ if the dark matter structure is suppressed on small scales, either by baryonic effects or by a model of dark matter alternative to CDM, as shown in Figure~\ref{fig:dCkk_WF}.  Instead, however, one can employ an alternative approach to characterizing the Gaussian bias, which entails randomizing the phases of the Fourier transforms of the CMB maps, prior to reconstructing the projected dark matter, as done in~\cite{actlensing}.  This phase randomization destroys any non-Gaussian lensing correlation between modes, while preserving the Gaussian bias term we want to subtract. We leave to future work the demonstration of this technique using high-resolution simulations.\\

{\it{Foregrounds:}} As discussed above, the SNR is largest when using CMB temperature maps, as opposed to polarization maps, to reconstruct the projected dark matter potential on multipoles of $L\sim 30,000$.  As a result, one needs to pay special attention to the impact of astrophysical foregrounds, such as the thermal Sunyaev-Zel'dovich (tSZ) signal from galaxy clusters, the microwave emission from AGN and star-forming galaxies (the latter of which is known as the Cosmic Infrared Background (CIB)), and the kinetic SZ (kSZ) effect from the velocity field of the dark matter~\cite[e.g.,][]{Sehgal2010}.  These foregrounds are significant in CMB temperature maps, whereas they are minimal in polarization maps.  They also contribute a non-Gaussian signal that can bias the lensing power spectrum~\cite{vanEngelen2014}.  In addition, there are still uncertainties in the Galactic foreground behavior at small scales~\cite{Fantaye2012}.  

Fortunately, there are a few paths one can take to foreground clean: \\

$\bullet$ Deproject foregrounds in the gradient-leg: One can use a recently proposed method to remove foreground biases by using foreground-cleaned CMB maps in the ``gradient leg'' of our lensing estimator~\cite{MH2018}.  Such foreground-cleaned gradient maps could be provided by an experiment like the Simons Observatory, which will have six frequency channels spanning 30 to 300 GHz~\cite{SO2018}.  As~\cite{MH2018} discuss, one can explicitly deproject foregrounds that have a known frequency dependence, such as the tSZ and CIB, from the gradient maps by using the constrained ILC formalism~\cite{CILC}. They show that this results in negligible foreground-bias to the lensing convergence map.  Explicitly deprojecting both the tSZ and CIB simultaneously from such upcoming lower-resolution temperature maps will not appreciably increase the noise in the gradient leg of the lensing estimator since the noise levels even after deprojection are expected to be well below the CMB signal for all $\ell$'s below 2000~\cite{SO2018}; thus the gradient maps are cosmic variance limited even after aggressive foreground cleaning.  

$\bullet$ Filter out all scales with $\ell < 5000$ in the non-gradient leg: Foregrounds in the non-gradient, small-scale leg will still add variance to the lensing reconstruction, however, one can filter out all scales with $\ell < 5000$ in the non-gradient leg with negligible loss of SNR, which removes the bulk of the tSZ signal and the clustering of galaxies.  After this filtering, the residual foregrounds in the non-gradient leg will consists mainly of Poisson distributed point sources and the kSZ.

$\bullet$ Remove Poisson point sources by template subtraction: Given the high-resolution (10 to 20 arcseconds) and low-noise (0.1$\mu$K-arcmin) of the non-gradient leg CMB map, one can identify individual point sources down to a low flux limit, measure their flux, and template-subtract them from the CMB maps using knowledge of the beam shape.  Template subtraction is effective because one does not cut holes in the CMB map or inpaint sources, both of which disrupt the resulting convergence map at the location of the sources.  With template-subtraction, the CMB under the sources remains intact.  Since CIB sources are the most dominant foreground at these small scales, we examine them in more detail.  The $1\sigma$ confusion limit for CIB sources at 350 GHz, given a 50-meter single dish, is about 0.03 mJy~\cite{Lagache2018}.  Thus a source detected at $5\sigma$ would have a flux of about 0.15 mJy at 350 GHz.  Template-subtracting these sources, after extrapolating their fluxes from 350 to 150 GHz, gives a flux cut of 0.016 mJy at 150 GHz.  Using the simulations of~\cite{Sehgal2010}, we find that this flux cut at 150 GHz lowers the CIB power at $\ell = 20,000$ by almost 5 orders of magnitude.  This makes the CIB power lower than the lensed CMB power at these high multipoles.  
 
$\bullet$ Use a shear-only reconstruction estimator: In addition to the methods described above, one can also use a novel technique of estimating the lensing convergence power spectrum from shear-only reconstructions~\cite{SchaanFerraro2018}. This method removes biases due to foregrounds such as the tSZ, CIB, and even the kSZ effect, the latter of which is naively hard to remove because it has no frequency dependence~\cite{FerraroHill2018}.  The shear-only estimator has a reduction in the SNR compared to the optimal quadratic estimator of about a factor of two.  Since the estimator employed in this work, which follows~\cite{hudedeovale2007}, has a larger reduction in SNR compared to the optimal quadratic estimator, the SNR can be expected to increase when using the shear-only estimator, which folds in small-scale information in both of its CMB map legs.  The maps that go into the shear-only estimator can first be foreground cleaned by 1.)~explicit deprojection using multi-frequency observations, 2.)~removing scales below $\ell < 5000$, and 3.)~template-subtraction of all sources with a flux above about 0.02 mJy. Then the shear-only reconstruction can be applied on the resulting maps.\\  

{\it{Other Mode-coupling Signals:}} When CMB instrumental noise levels are as low as about $0.25 \mu$K-arcmin, a potentially limiting factor to the perfect reconstruction of the lensing field is that the lensing field also has some rotation.  This rotation arises because there is more than one lens plane and because the lensing is not perfectly weak, making the first-order Born approximation inexact~\cite[e.g.,][]{HirataSeljak2003,Pratten2016,Lewis2017,Fabbian2018,Beck2018}.  The Born approximation is when the lensing deflections are computed along the unperturbed photon path, as opposed to the perturbed photon path. Since the lensing field has some rotation, which we call ``curl'' mode-coupling, the concern is that this mode-coupling will ``leak'' into the lensing convergence mode coupling we are interested in isolating.  This effect has been shown to be negligible for CMB-S4-type sensitivities~\cite{Beck2018}, however could be significant for the sensitivities and scales discussed here.  

However, we can isolate the convergence and ``curl'' mode couplings using a “bias-hardening” technique proposed by ~\cite{Osborne2014}. Referring to Eqs. 21 and 22 of that paper, one replaces the $S^2$ term in those equations with the rotation component.  The idea with bias-hardening is that the convergence and rotation do not need to be independent.  However, as long as they are not completely degenerate, which they are not, one can create new “bias-hardened” estimators for each, where the off-diagonal correlations are zero (by diagonalizing the response matrix of known response coefficients).  Then, with these two “bias-hardened” measurements, one has a system of two-equations and two unknowns (Eq 22), and can solve for the convergence piece alone. 

We also note that post-Born corrections add extra power both to the lensing power spectrum and to the lensed CMB power spectra, on the level of a percent at these high multipoles~\cite{Pratten2016,Fabbian2018}. We propagate a $1\%$ increase in the lensed CMB TT power spectrum for all $\ell > 4000$ through our signal-to-noise calculations, keeping the theory CMB spectrum in the lensing filteres unchanged, in order to simulate our potential ignorance of higher-order effects in the CMB spectrum.  This results in a change of $0.002\%$ to the SNRs. \\

{\it{Non-Gaussianity of the Matter Power Spectrum:}}
The simulation-based covariance matrix we employ in these forecasts, and discuss in detail in the appendix, capture higher-order lensing corrections arising from Gaussian realizations of the CMB lensing convergence field.  However, we note that there exist non-Gaussian fluctuations of the matter power spectrum that are significant on these small scales~\cite{Bohm2016,Bohm2018,Beck2018}. A measurement of the CMB lensing power spectrum as proposed above will naturally be sensitive to this non-Gaussian signal.  

To assess the impact of these non-Gaussian convergence map fluctuations on the SNR and to quantify any $N^{3/2}$ bias, we employ simulations by~\cite{Liu2018} which include CMB lensing convergence maps where the lensing was done by ray-tracing through structure in an N-body simulation.  We note that while the convergence maps from~\cite{Liu2018} are the best publicly available for these tests, limitations of N-body resolution and the ray-tracing procedure in those simulations might not allow all non-Gaussian effects to be captured at small scales.  We follow the procedure outlined in the appendix to create the covariance matrix, but instead use 1,000 realizations of these new simulations, each covering about 12 square degrees. We find that compared to the SNRs from Gaussian convergence simulations, the SNRs from N-body convergence simulations are degraded by 6\% for the 18'' resolution case and by 9\% for the 9.5'' resolution case, both assuming 0.1$\mu$K-arcmin noise. From the comparison of lensing auto power spectra, we observe an average $N^{3/2}$ bias between $L=5,000$ and $30,000$ of 2\%, with a maximum bias in this range of 10\% (between $L=25,000$ and $30,000$ where the contribution to the SNR is low).  Part of the reason that the $N^{3/2}$ bias is so low out to these large lensing multipoles is because we have employed a gradient cut in one of our lensing estimator legs. Thus the $N^{3/2}$ bias appears only via secondary contractions in the estimator.  We make no attempt to remove this small $N^{3/2}$ bias from the measurement since it contains cosmological information and should be forward modeled in any theory prediction.  \\

\section{Discussion}
\label{sec:discussion}

We have shown that very high-resolution CMB lensing measurements have the statistical potential to provide high-significance measurements of the small-scale matter power spectrum.  We have also identified the primary systematic effects of concern for this measurement, as well as ways to mitigate them.  A more complete study of systematic effects applicable to this technique is left to future work.  

A similar measurement of the small-scale matter power spectrum might be possible using galaxy shear information either through cosmic shear measurements or cross-correlations between CMB lensing, galaxy shear, and galaxy counts. Since for the same angular scale on the sky, sources and tracers at lower redshifts probe smaller scales in the 3-dimensional matter power spectrum, and since existing and planned galaxy surveys like LSST already have the resolving capability for $L\sim 30,000$, such measurements could have more statistical significance than the CMB lensing approach presented here. However, one complication is correlated modes on these small scales arising from, for example, point-spread-function uncertainties.  In the case of CMB lensing, realization-dependent $N_0$ subtraction, as we discuss in the appendix, minimizes the correlation between modes.  In addition, complications arising from imperfect shear measurement, blending of galaxies, and photometric redshift uncertainty make such measurements challenging from a systematics perspective. We leave further consideration of the potential for small-scale galaxy lensing to future work.

Realizing the potential of high-resolution CMB lensing would yield the advantage that small-scale structure would be probed i)~directly via gravitational lensing, ii)~at relatively high-redshifts ($z\approx 1-3$) where a deviation from CDM is clearer, iii) with high statistical significance, and iv)~and with potentially minimal systematic uncertainty.  With such a measurement, one could robustly determine whether structure is suppressed at small scales, in deviation from the dark-matter-only CDM prediction.  If such a deviation is found, one could also potentially distinguish between a baryon-induced suppression or a suppression arising due to the particle nature of dark matter.  The instrumentation required for such a measurement, both the needed dish size and camera sensitivity, are within reach of a future generation of ground-based CMB experiments.  High-resolution CMB lensing may provide a powerful and robust approach to measuring the small-scale matter power spectrum, informing both baryonic physics and the nature of dark matter.

\begin{acknowledgments}
The authors thank Chi-Ting Chiang, Rouven Essig, Simone Ferraro, Sunil Golwala, Dongwon Han, Renee Hlozek, Gil Holder, Manoj Kaplinghat, Marilena Loverde, David Marsh, Joel Meyers, Emmanuel Schaan, Blake Sherwin, Kendrick Smith, Sean Tulin, Alexander van Engelen, and Grant Wilson for useful discussions.  We thank Jia Liu for providing us with CMB lensing convergence simulations where the lensing was done by ray-tracing through an N-body simulation.  We also thank the referee for insightful comments that improved this work.  HNN acknowledges support from the URECA (Undergraduate Research \& Creative Activities) Summer Program, and NS acknowledges support from NSF grant number 1513618.
\end{acknowledgments}

\appendix
\section{Simulation-based Covariance Matrix}

To accurately forecast the statistical significance of a high-$L$ measurement of the lensing power spectrum with a reliable covariance matrix, we perform simulations of the reconstruction process using the estimator given in~\cite{hudedeovale2007}. The covariance matrix assumes the fiducial CDM cosmology described in the main text.

\begin{figure}[t]
\centering
\includegraphics[width=0.95\columnwidth]{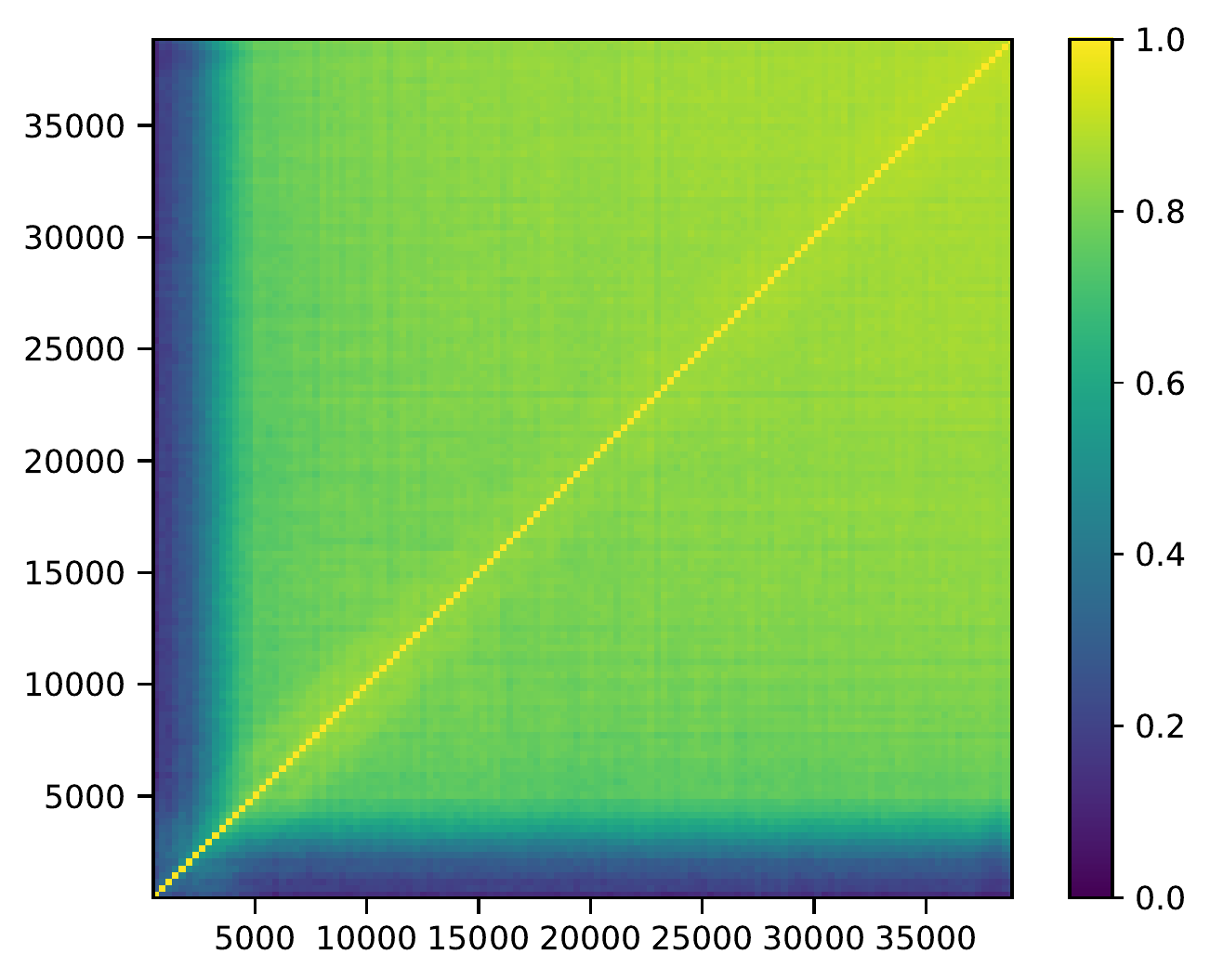}
\includegraphics[width=0.95\columnwidth]{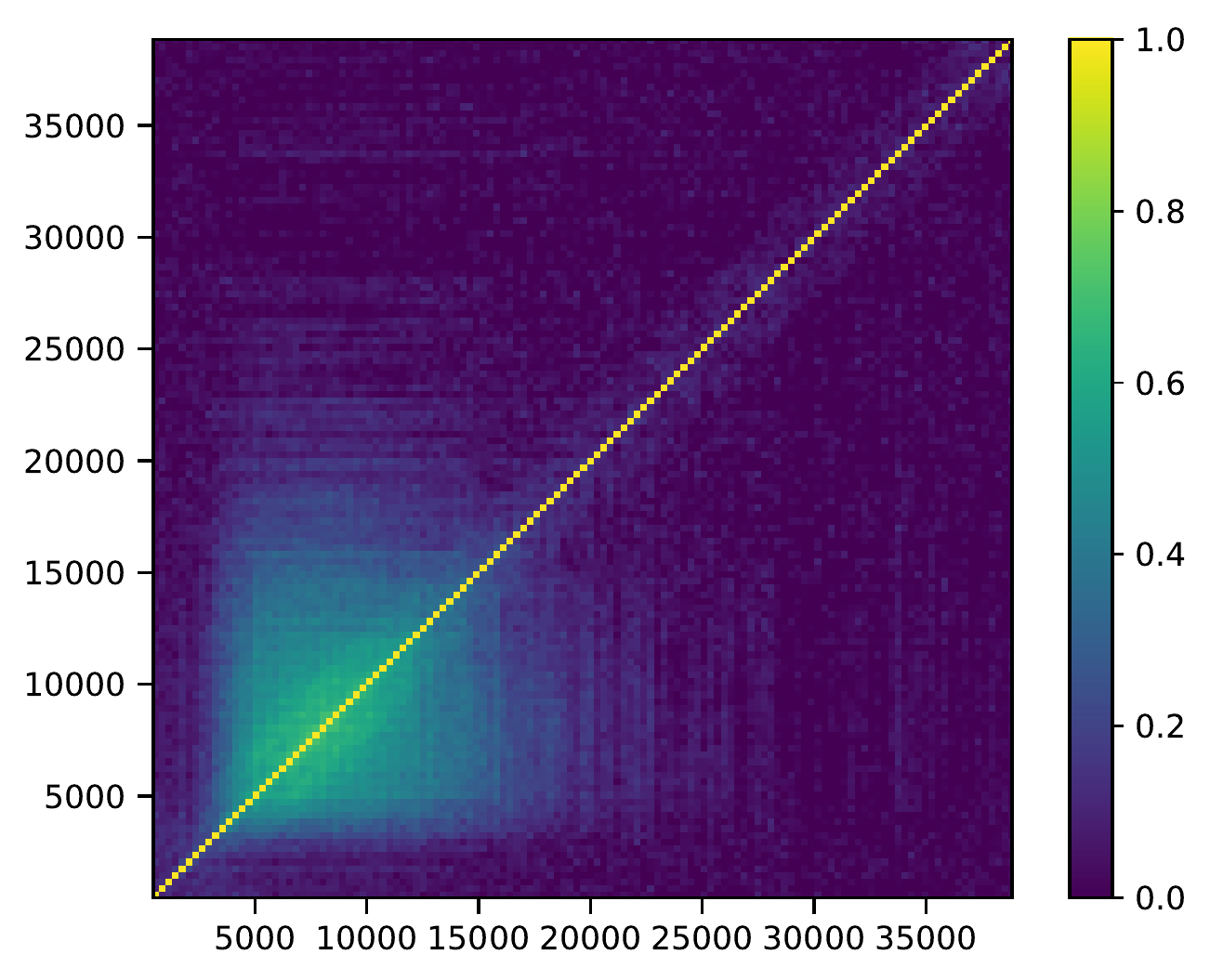}
\caption{Bin-to-bin correlation coefficients of the lensing power spectrum for an experiment with 18" beam and 0.5~$\mu$K-arcmin white noise. {\it Top panel:} Coefficients when no $N_0$ is subtracted from the naive power spectrum estimate (or equivalently when the same $N_0$ is subtracted from each simulation realization). {\it Bottom panel:} Correlation coefficients obtained when a realization-dependent $N_0$ subtraction is done for each simulation.}
\label{fig:covMat}
\end{figure} 

Periodic Gaussian random field realizations of the unlensed CMB power spectrum are prepared on patches with $2048\times 2048$ pixels and a pixel width of 0.05 arcminutes. These are lensed by interpolating pixel displacements (with 5th order spline interpolation) obtained from the appropriate transform of periodic Gaussian random realizations of CMB lensing convergence fields with a lensing power spectrum given by our fiducial CDM cosmology. The lensed CMB is then beam convolved and a random realization of the appropriate instrumental white noise is added. For each experimental configuration, 1000 simulations, each with independent realizations of unlensed CMB, lensing convergence, and instrumental noise are prepared. These are then downsampled in Fourier space to our analysis resolution of $1024\times 1024$ pixels and a pixel width of 0.1 arcminutes. The Fourier-space downsampling, achieved by trimming the map in Fourier space so as to cut modes below the target pixel scale, circumvents the need to account for a pixel window function and speeds up lensing reconstruction while preserving the CMB modes of interest.
 
\begin{figure}[t]
\centering
\includegraphics[width=0.5\textwidth]{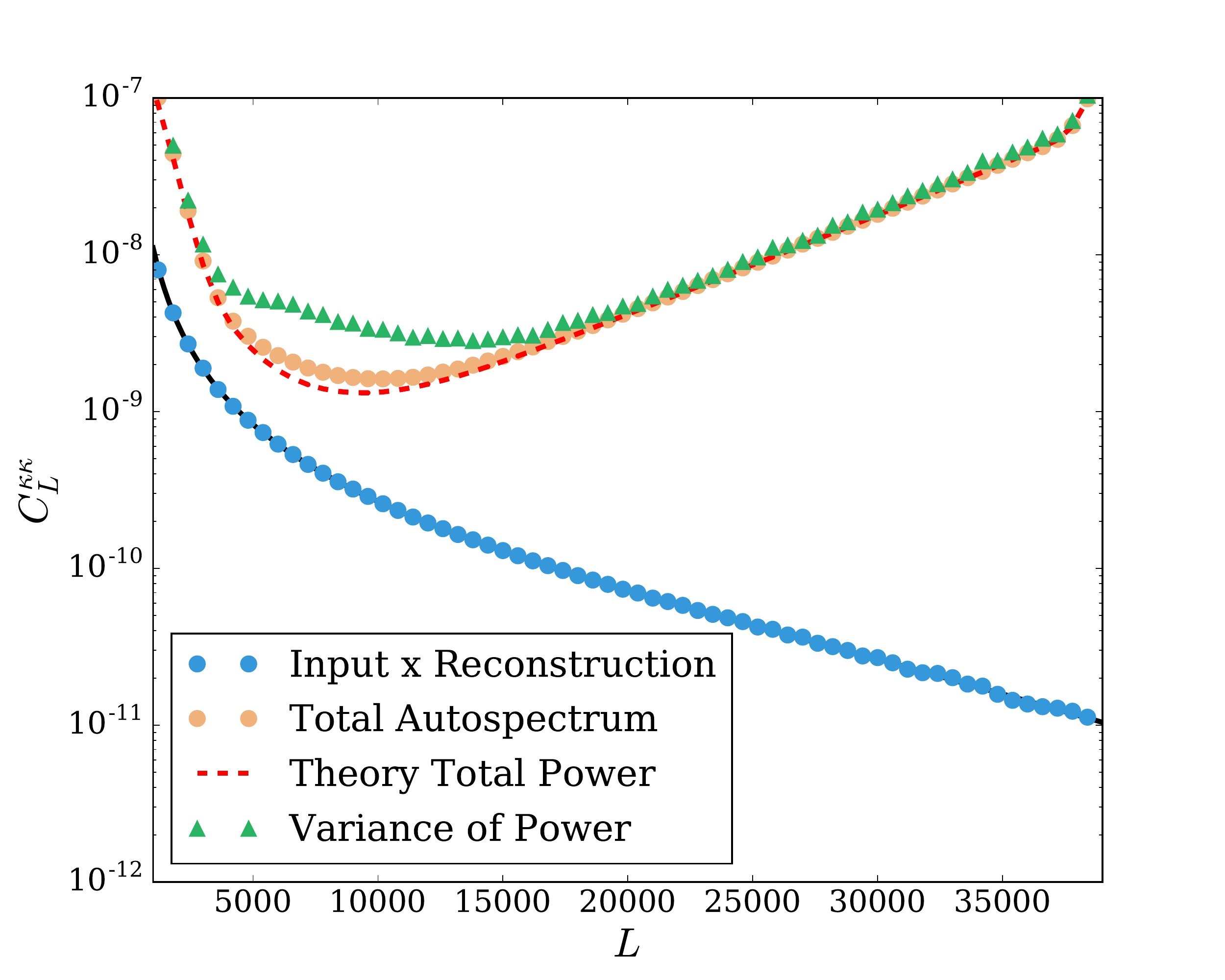}
\caption{Lensing bandpowers from simulated lensing reconstructions. The blue circles are the binned cross-spectrum of the input lensing convergence map and the corresponding reconstruction. The black solid line is the theory spectrum used to generate the input lensing maps. The red dashed line is the predicted total power in the map (i.e. the sum of $C_L^{\kappa \kappa}$ and a theory estimate of $N_L^{\kappa\kappa}$ from Eq.~\ref{eq:NL}). The orange circles are the binned, simulation-based lensing power spectrum with no $N_0$ bias subtraction. The green triangles are the variance in the power obtained directly from the diagonal terms of the covariance matrix of the power spectra of the reconstructions after a realization-dependent $N_0$ subtraction has been performed.}
\label{fig:covMatDiagonals}
\end{figure} 
 
Since most of the signal-to-noise is in the TTTT channel for the CMB lensing auto-spectrum, for each simulated CMB map, $T_i$, we obtain a convergence estimate $\hat{\kappa}_i$ using the TT estimator of~\cite{hudedeovale2007}, with the gradient leg low-pass filtered to remove $\ell<2000$. This low-pass filtering of the gradient results in a loss of signal-to-noise only at low lensing multipoles $L$, which are irrelevant to our analysis.  However, this cut is essential to avoid the $N_2$ bias~\cite{Hanson2011}, which appears both as a multiplicative bias in the lensing convergence map and as bias in the lensing power spectrum. We first construct the naive power spectrum $\langle\hat{\kappa}_i(L)\hat{\kappa}^*_i(L)\rangle$,
and then subtract from it the $N_0$ bias using Eq.~17 of~\cite{huokamoto2002}, but with the CMB power spectra in the integral scaled by the ratio of the actual total power spectrum in the CMB map to the fiducial total power spectrum used to generate the simulations. This procedure mimics the realization-dependent noise bias subtraction from~\cite{Namikawa2013} and effectively accounts for variations in the noise bias among different simulations.

Subtracting the $N_0$ bias in this way, which would be done in a realistic analysis of CMB data, has two advantages. First it is robust to mismatches between the data and the simulations used to calculate the $N_0$ bias. Second, and more important to this work, this subtraction improves the covariance properties of the noise matrix. In the top panel of Figure~\ref{fig:covMat}, we show the correlation coefficients of the band-powers of the CMB lensing reconstruction if no $N_0$ bias were subtracted from the lensing power spectra (or equivalently, if a mean simulated $N_0$ bias that did not change from realization to realization were subtracted). Bandpowers with $L>5000$ become almost completely correlated, signaling almost complete loss of information above this scale. This correlation is expected in the high-$L$ limit of the quadratic estimator (see e.g.~\cite{FerraroSherwin2017,Horowitz2017}). However, as seen in the bottom panel of Figure~\ref{fig:covMat}, the correlation coefficients are much smaller if the $N_0$ bias is subtracted in the realization-dependent manner described above. Here, bandpowers are now only significantly correlated in a relatively small $L$-range of $5,000<L<12,000$.

In Figure~\ref{fig:covMatDiagonals}, we show the diagonal terms of the noise covariance matrix after $N_0$ bias subtraction, i.e., the variance of the CMB lensing bandpowers. We find that the diagonal variance from simulations is larger than that predicted from the $N_L^{\kappa\kappa}$ of Eq.~\ref{eq:NL}, and cannot be accounted for by the next-order $N_1$ contribution, which we calculate using simulations following~\cite{Sherwin2017}. The excess variance is likely due to higher-order lensing corrections captured by our simulations.  While the residual covariance and excess in diagonal variance might be mitigated by improved lensing estimators, our forecasts in Table~\ref{tab:sn} use the full covariance matrix from simulations, with realization-dependent $N_0$ subtraction, described above.

\end{document}